\documentstyle{article}
\begin{document}

\title{ The cyclotron emission of anisotropic electrons in the X-ray pulsars}
 \author{ A.N.Baushev\thanks
 {Space Research Institute, Profsoyuznaya,
 84/32, Moscow 117810, Russia}
 and G.S.Bisnovatyi-Kogan\thanks
 {Space Research Institute, Profsoyuznaya,
 84/32, Moscow 117810, Russia, E-mail: gkogan@mx.iki.rssi.ru}}

 \date{}
 \maketitle

\begin{abstract}
The spectrum of cyclotron radiation is calculated for anisotropically
distributed relativistic electrons with a nonrelativistic velocity
scattering across the magnetic field. It is shown that if such electrons
are responsible for a formation of the "cyclotron" line in the spectrum
of Her X-1, then the value of its magnetic field $(3-6)\cdot 10^{10}$ Gs
following from this interpretation is in a good agreement with some other
observations and theoretical estimations. Observations of a time dependence
of the energy of this "cyclotron" line in the spectra of several X-ray
pulsars is explained by a variability of the average longitude energy of
the electrons, decreasing with increasing of the liminocity due to
radiational braking of the accretion flow.
\end{abstract}

\section{Introduction.}

  There are a lot of observations of the cyclotron
resonance structure in the X-ray spectrum of pulsar Her X-1
\cite{a,b,c,d,e,f}
at 39-58 KeV (see Table).
This singularity is interpreted as a cyclotron line,
and the magnetic field intensity was usually calculated from the
non-relativistic formula

                       $$H=\frac{m c \omega}{e},  \eqno(1)$$
where $\omega$ is the cycle frequency of the photons, $m$ is the mass of
the electron. It was obtained
to be of the order of $(3-5)\cdot 10^{12}$~Gs.
But as large as this value comes into conflict with some theoretical
reasonings like interpretation of the observations of pulsar spin
acceleration \cite{h}, condition for the transparency for the outgoing of the
directed radiation \cite{i,u},
consideration of the interrelation between radio and X-ray pulsars
\cite{j}, simulation of the 35-days cycle variability \cite{g},
see also \cite{lip}

It seems likely that the reason of this conflict is an unsuitability of the
non-relativistic formula in this case. According to \cite{k}, the temperature
of the electrons emitting a cyclotron line could be $\sim 10^{11}K$,
and therefore they are ultrarelativistic.
By this means the mean energy of the cyclotron line is broadened and shifted
relativistically by a factor of $\gamma \simeq \frac{kT}{mc^2}$. In this
article the spectrum profile of the cyclotron line is calculated for
various electrons distributions. Furthermore, the model of the hot spot
 on the pulsar is considered and it is shown that the overall observed
X-ray spectrum (from 0.2 to 120 KeV) can arise under the fields near
the pulsar surface ($\approx 5\cdot 10^{10}$~Gs) which are well below
 then those, obtaind from (1).

\section{The cyclotron radiation of the anisotropic relativistic electrons}

According to \cite{l,i}, in the magnetic field near the pulsar the cross
component of a momentum emits rapidly, while the parallel component remains
constant.
Hence the momentum distribution of the electrons is
anisotropic
 $$p_\perp^2\ll p_\parallel^2,      \eqno(2)$$
 where $p_\perp\ll mc$,
$p_\parallel\gg mc$.
In this article we assume for simplicity that the transverse electron
distribution is two-dimensional Maxwellian
$$
  dn=\frac{N}{T_1}{\exp \left( -\frac{m u^{2}}{2T_1} \right)}\,
  d\frac{m u^{2}}{2},
         \eqno(3)
$$
where  $T_1 \ll mc^2$.
Let us calculate the cyclotron emission of N such particles that move
at a rate  $V$ along the magnetic field.

For a single particle, having the transverse velocity $u$, we find \cite{m}:
$$
j(\theta)=\frac{e^4 H^2 u^2 (1-\frac{V^2}{c^2})^2
[(1+\cos\theta)(1+\frac{V^2}{c^2})-4\frac{V}{c} \cos\theta]}
{8 \pi c^5 m^2 (1-\frac{V}{c} \cos\theta)^5},          \eqno(4)
$$
where $\theta$ is an observational angle in a laboratory frame of reference.
Integrating over the distribution (3), we obtain for $N$ particles:
$$
J(\theta)=\int j(\theta) dn=N \frac{e^4 H^2 T_1 (1-\frac{V^2}{c^2})^2
[(1+\cos\theta)(1+\frac{V^2}{c^2})-4\frac{V}{c} \cos\theta]}
{4 \pi c^5 m^3 (1-\frac{V}{c} \cos\theta)^5}.          \eqno(5)
$$
For the spectrum we find:
$$
\omega (\theta)=\omega_H \frac{\sqrt {1-\frac{V^2}{c^2}}}
{1-\frac{V}{c}\cos\theta},\quad
\omega_H=\frac{eH}{m c}.              \eqno(6)
$$
When $V\simeq c$ the cyclotron radiation is highly directed and diagram
has a pencil beam along V ($\theta \simeq 0$). Under these conditions
 $(\theta=0, V\simeq c)$ we obtain from (5),(6):
$$
J(0)=\frac{2 N e^4 H^2 T_1}{\pi c^5 m^3 (1-\frac{V}{c})},  \eqno(7)
$$
$$
\omega(0)=\omega_H \sqrt \frac{1+\frac{V}{c}}{1-\frac{V}{c}}
\approx 2 \omega_H\frac{E_\parallel}{m_e c^2},   \eqno(8)
$$
what gives
$$
1-\frac{V}{c}=\frac{2 \omega_H^2}{\omega^2}.    \eqno(9)
$$
Let us consider the parallel momentum distribution of the electrons as:
$$
dn=f(p_\parallel)\,dp_\parallel.                 \eqno(10)
$$
Substituting of $dn$ for $N$ and using

$$
p_\parallel=\frac{mc}{2}\frac{\omega}{\omega_H}\, ;\quad
 1-\frac{V}{c}=\frac{2 \omega_H^2}{\omega^2},    \eqno(11)
$$
we obtain for the spectral density:

$$
J_\omega=\frac{e^2 T}{2 \pi c^2 \omega_H} \omega^2
 f \left( -\frac{m c }{2}\frac{\omega}{\omega_H}\right )\,d\omega.  \eqno(12)
$$
Let us consider two important cases. When $f$ is a relativistic Maxwell:
$$
f=\frac{n_0 c}{T_2} \exp\left( -\frac{p_\parallel c}{T_2}\right )\,  , \quad
T_2\gg m c^2\gg T_1, \eqno(13)
$$
where $n_0$ is a number of emitting electrons. Then the spectrum is:
$$
J_\omega=\frac{n_0 e^2}{2 \pi c \omega_H}\frac{T}{T_2} \omega^2
 \exp\left( -\frac{m c^2 \omega}{2 \omega_H T_2}\right )
\,d\omega.  \eqno(14)
$$
This spectrum has a single maximum at
$$
\frac{\omega}{\omega_H}=\frac{4 T_2}{m c^2}.  \eqno(15)
$$
In the second case consider the function $f$ as
$$
f=\frac{n_0 }{\sqrt {\pi} \sigma}
\exp\left( -\frac{{(p_\parallel-a)}^2}{\sigma^2}\right ).
\eqno(16)
$$
The spectrum of radiation is
$$
J_\omega=\frac{n_0 e^2}{2 \pi c^2 \omega_H} \omega^2
 \exp\left( -\frac{(\frac{m c}{2}\frac{\omega}{\omega_H }-a)^2}
 {\sigma^2}\right )
\,d\omega.  \eqno(17)
$$
When $\sigma \ll a$ this spectrum has a single maximum at
$$
\omega \simeq \frac{2 a}{m c} \omega_H. \eqno(18)
$$
Notice that in all cases the maximum is shifted to
$$
\frac{\omega}{\omega_H} \sim \frac{\bar E_e}{m c^2}. \eqno(19)
$$
It is a common property of relativistic cyclotron line, that is independent
of the particular form of $f$.
We had approximated experimental spectrum taken from \cite{o} (solid line)
by (14),(17) (dot line in fig. 1 and 2), and spectrum \cite{p} by (17) only
(fig. 3). Setting (in accordance with  \cite{k}) the longitude
electron temperature as $\sim 2 \cdot 10^{11}$~K, that is
$T_2= 2\cdot 10^{11}$~K and $a=7 \cdot 10^{-4}$~${\rm\frac{eV \cdot s }{cm}}$,
we obtain for the magnetic field strength $B=4\cdot 10^{10}$~Gs,
$8\cdot 10^{10}$~Gs, and $4\cdot 10^{10}$~Gs respectively.
Here we estimate the spectral form of the cyclotron line averaged over the pulsar period,
supposing uniform distribution of $f(p\parallel)$ over the polar cap.
In this model the beam of the cyclotron feature is determined by the
number distribution of the emitting relativistic electrons,
moving predominantly
along the magnetic field, over the polar cap.

\section{Model of the X-ray spectrum of Her X-1.}

In order to obtain the whole experimental spectrum of the Her X-1
the following model of the hot spot is considered.
A collisionless shock wave is generated in the accretion stream
nearby the surface on the magnetic pole of a neutron star. In it`s front the
ultrarelativistic electrons are generated. It's worth mentioning that
according to \cite{i,l} these electrons when generated can possess
only small pitch-angle values and the condition (2) is fullfilled
automatically. Under the shock
there is a hot turbulent zone with a temperature $T_e$,
  and optical depth $\tau_e$, and under this zone a heated spot on the surface
of the neutron star with a smaller temperature is situated.

The whole X-ray spectrum of pulsar Her X-1 is represented on fig.3 by the solid
line. It was taken from \cite{p}. There are three main regions in it:
a quasi-Plankian spectrum between 0,3 and 0,6 KeV, that is generated
near the magnetosphere of the X-ray pulsar; power-law spectrum $(0.6\div
20)$~KeV with a rapid decrease at 20 KeV, and the cyclotron feature.

The power-law spectrum area appears as follows.
A surface emits the black-body spectrum with
a temperature $T_s$. Travelling through the turbulent zone this radiation is
comptonized. This comptonized spectrum has been calculated according to
\cite{s}. Setting the neutron star radius equal to 10 km, distance from the
X-ray pulsar 6 Kps, hot spot area $S=2 \cdot 10^{12}$~cm$^2$,
we have found the best
approximation conditions at $T_s=1$~KeV, $T_e=8$~KeV, $\tau_e=14$.
The best approximation of the X-ray spectrum of the pulsar Her X-1 is
represented in fig.3 by the dot line. It agrees nicely with the experimental
curve.

\section{Discussion.}

The observation of the variability of the cyclotron line
is reported in \cite{n}. Ginga detected the changes of the cyclotron energies
from 4 pulsars. The change is as much as 40 \% in the case of 4U 0115+63.
Larger luminosity of the source corresponds to smaller average energy of the
cyclotron feature. These changes might be easily explained in our model.
The velocity of the accretion flow decreases with increasing of the
pulsar`s
 luminosity   because localy the luminosity is close to the Eddington limit.
As a result the shock wave intensity drops as well as the energy of the
ultrarelativistic electrons in it`s front. The cyclotron energy decreases in
accordance with (19).

\section{Conclusion.}

X-ray pulsar Her X-1 is one of the most interesting and most investigated
of this kind of objects. In the spectra of the other pulsars there is also
observed the cyclotron line, but these observations were less reliable.
In all of these cases the magnetic field intensity turned out to be too large,
if it is calculated according to the non-relativistic formula.
The way to overcome this difficulty is proposed in this article.
So the relativistic formula for the cyclotron line yields for
the magnetic fields the value that is consistent with other
observational data and many theoretical
estimates.

\begin{figure}
\begin{tabular}{|l|c|c|c|}
\hline
\hline
Date&Article&$\omega_{\rm max}$(KeV)&Width(KeV)\\
\hline
1978,May&\cite{b}&58&$11^{+26}_{-11}$\\
\hline
1977,Sep.&\cite{c}&51&$21^{+9}_{-7}$\\
\hline
1978,Feb.&\cite{d}&48&$28\pm 7$\\
\hline
1980,Apr.&\cite{d}&54&$11^{+14}_{-11}$\\
\hline
1980,May.&\cite{e}&49.5&$18^{+6}_{-3}$\\
\hline
1980,Sep.&\cite{a}&39&$27^{+21}_{-20}$\\
\hline
\end{tabular}
\end{figure}

\vfill \eject

\centerline{\bf Figure captions}
\medskip
{\bf fig.1} Comparision of the observational and computational spectra of the
cyclotron line. The solid curve is the observational results taken from
\cite{o}, the dot curve is the approximation by the comptonized 
spectrum, and feature (14) with
  $T_2= 2\cdot
10^{11}$~K.

  \medskip
 {\bf fig.2}   Comparision of the observational and computational spectra of the
cyclotron line. The solid curve is the observational results taken from
\cite{o}, the dot curve is the approximation by the comptonized 
spectrum, and feature (17) with
 $a=7 \cdot 10^{-4}$~${\rm\frac{eV \cdot s }{cm}}$,
 $\sigma=2 \cdot 10^{-4}$~${\rm\frac{eV \cdot s }{cm}}$.

 \medskip
  {\bf fig.3} Comparision of the observational and computational X-ray spectra
of Her X-1. The solid curve is the observational results taken from \cite{p},
the dot curve is the approximation with 
$T_s=0.9$~KeV, $T_e=8$~KeV, $\tau_e=14$,
$a=7 \cdot 10^{-4}$~${\rm\frac{eV \cdot s }{cm}} $,
$\sigma=10^{-4}$~${\rm\frac{eV \cdot s }{cm}} $.

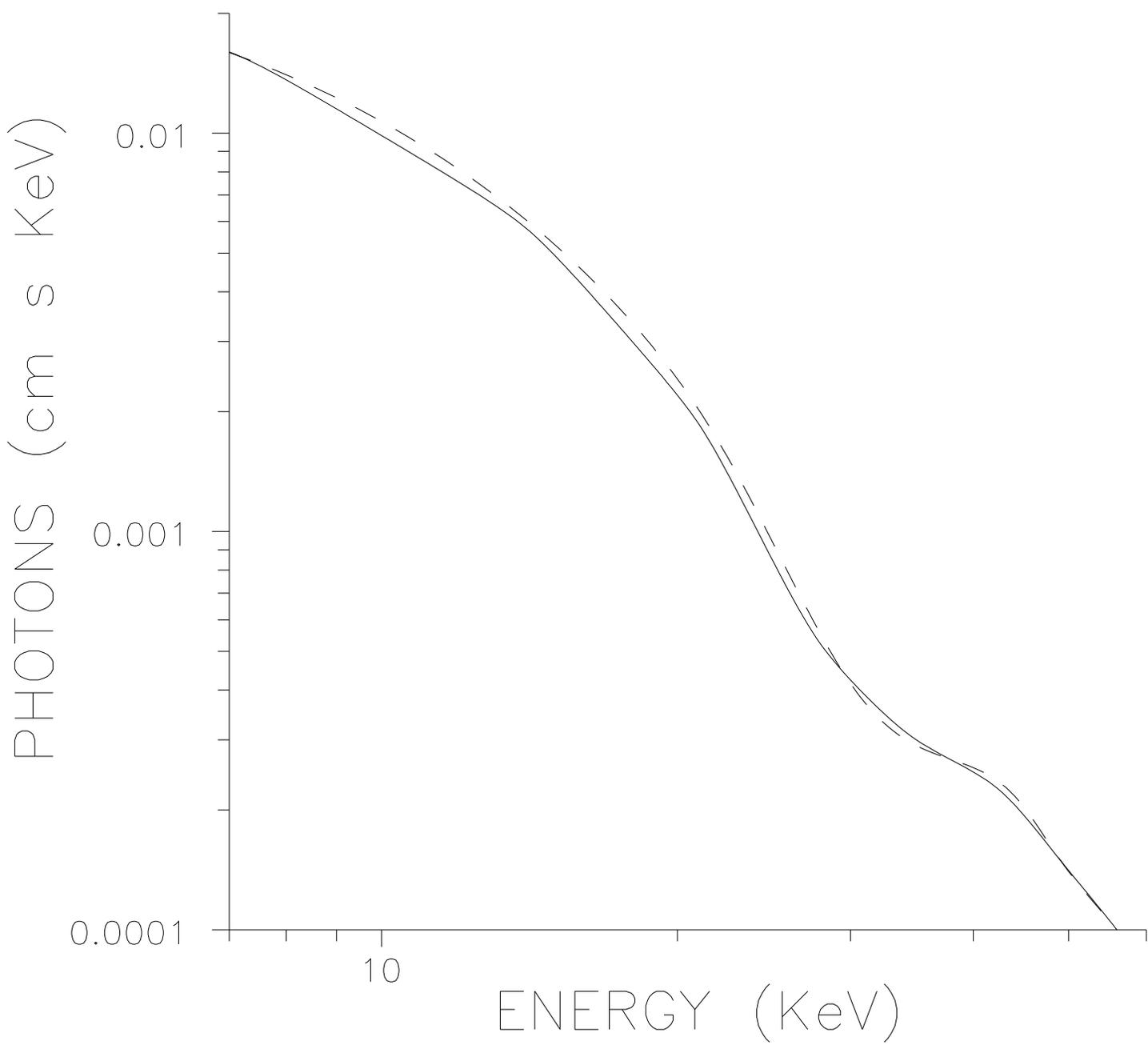
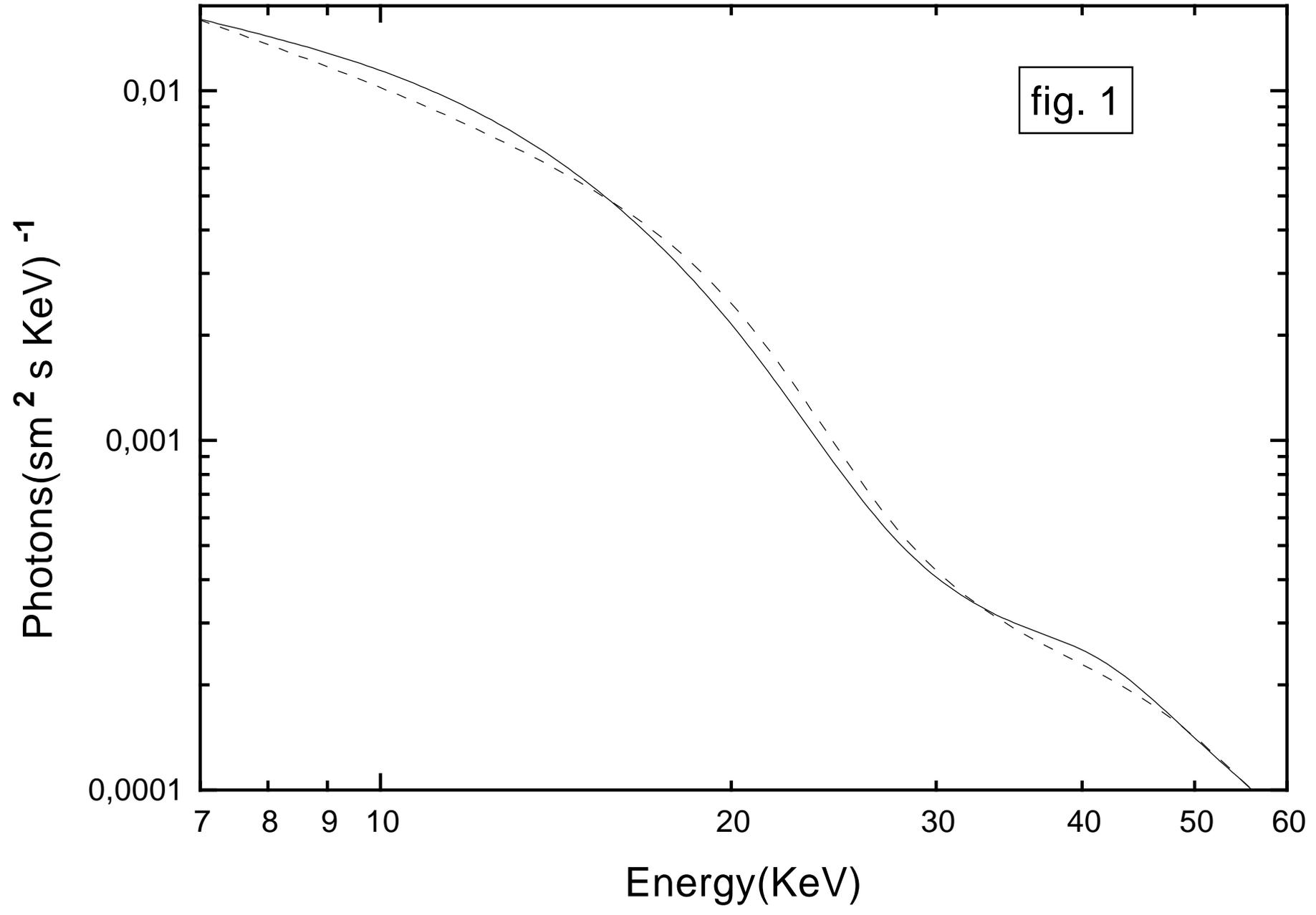
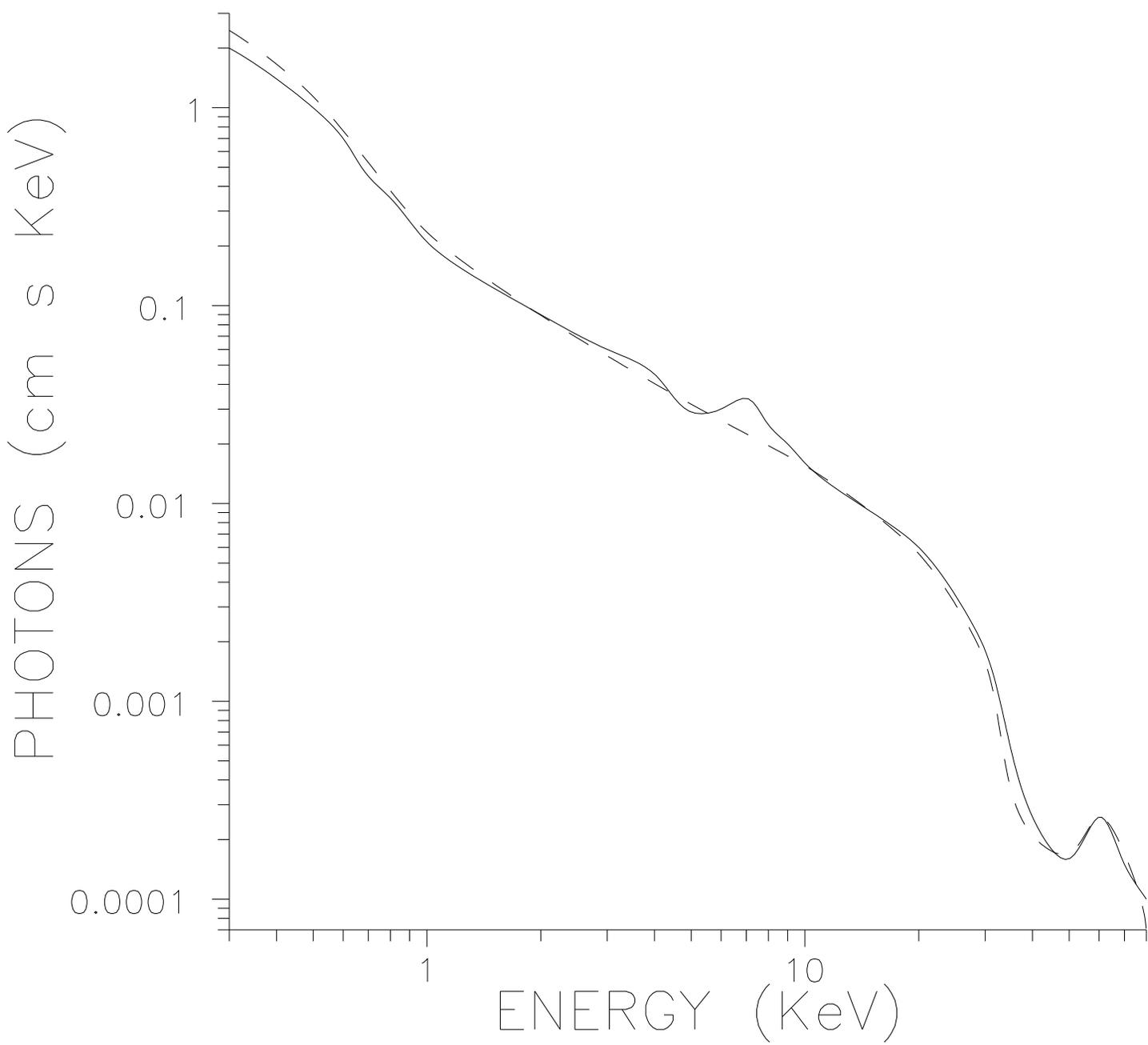
\end{document}